\documentstyle[aps,twocolumn,tighten,epsfig]{revtex}
\begin{document}

\draft
\title{Antibaryon density in the central rapidity region of a 
       heavy-ion collision}

\author{Jacek Dziarmaga$^1$
        and Mariusz Sadzikowski$^2$}
\address{ 1) Institute of Physics, Jagiellonian University,
          Reymonta 4, 30-059 Krak\'ow, Poland\\
          2) Institute of Nuclear Physics, Radzikowskiego 152,
             31-342 Krak\'ow, Poland}
\date{September 9, 1998}
\maketitle
\tighten

\begin{abstract}
We consider (anti-)baryons production in heavy ion collisions as production 
of topological defects during the chiral phase transition. Non-zero quark masses 
which explicitly break chiral symmetry supress the (anti-)baryon density. 
Hardly any (anti-)baryons will be produced in the central rapidity region of 
a heavy ion collision.
\end{abstract}
\vspace*{0.5cm}

   In the series of papers Ellis and Kowalski \cite{ellis1} and Ellis,
Kowalski and Heinz \cite{ellis2} considered baryon and antibaryon production
using a topological model inspired by the Skyrme picture of baryons as
topological defects in the quark-antiquark condensate \cite{skyrme,witten}.
The key idea is that the hadronization process is associated with spontaneous
chiral symmetry breaking. During these processes domains are formed in which
the order parameter, the quark condensate $\langle 0|\bar{q} q |0\rangle $,
takes the vacuum expectation value in each domain independently. Defects
may form if orientation of the order parameter in adjacent domains has a
topologically non-trivial configuration. These defects are interpreted as
baryons. Geometrical consideration in the case of heavy-ion collisions
predicts 0.17-0.35 baryons/fm$^3$ \cite{ellis2}.
   
   In this communication we would like to consider the same scenario under
the assumption that the chiral phase transition is described by the $O(4)$
sigma model nearby the critical temperature \cite{wilczek1,wilczek3,wilczek2}.
The massless two flavours case was considered by Gill \cite{gill} who used
Zurek scenario \cite{zurek} to estimate the density of topological defects
after a quench. However, it is known \cite{jacek} that explicit chiral
symmetry breaking through the non-zero quark masses leads to an important
modification of the Zurek picture. We will show that this effect results in
substantial suppression of the baryon production.

   In the Bjorken scenario \cite{bjorken} the quark gluon plasma comes to
thermal equilibrium at the temperature $T_I$ around $\tau_{I}=1\mbox{fm}$
after an ultrarelativistic heavy ion collision. As the ultrarelativistic
plasma is expanding, it cools down adiabatically according to

\begin{equation}\label{T}
T=T_I (\frac{\tau_{I}}{\tau})^{1/3} \;\;.
\end{equation}
$\tau=\sqrt{t^2-z^2}$ is the proper time. Quarks and gluons are the relevant
degrees of freedom as long as the temperature does not drop below $T_{conf}$,
where the confinement-deconfinement phase transition takes
place. Even above $T_{conf}$ the chiral fields can be defined and measured
with appropriate $<\bar{q}q>$ expectation values. One can measure e.g.
some pion fluctuations. However, the pions do not have any dynamics
of their own, they are mere phantoms which come out from appropriate
quark-antiquark correlations in the quark-gluon plasma. As a result
above $T_{conf}$ the chiral fields are {\it by definition} in thermal
equilibrium with the thermalized plasma.

   The situation changes at $T=T_{conf}$. The plasma disappears and the
chiral degrees of freedom become relevant. If confinement - deconfinement
phase transition is first order the takeover proceeds at
a constant temperature: latent heat is released and the system cools down
by adiabatic expansion. The chiral fields enter the $T<T_{conf}$ stage
in the state of thermal equilibrium and they begin to evolve without any
quark-gluon heat bath.

   The free evolution of the chiral fields is described by the linear
$O(4)$ sigma model with an explicit chiral symmetry breaking term

\begin{eqnarray}\label{model}
L &=& \frac{1}{2}\partial_{\mu}\phi^{a}\partial^{\mu}\phi^{a} - V \;\;,\nonumber\\
V &=& - \frac{A}{2} (\phi^{a}\phi^{a}) +
      \frac{\lambda}{4}(\phi^{a}\phi^{a})^2 -
      H\sigma  \;\;,
\end{eqnarray}
where the vector $\phi=(\sigma,\vec{\pi})$ is an $O(4)$ multiplet of real
scalar fields. The vector $\vec{\pi}$ represents the pion fields. The
typical zero temperature fits of some of the parameters are \cite{wilczek2}:

\begin{eqnarray}\label{constants}
&& \lambda=20 \;\;,     \nonumber\\
&& H=(119\;\mbox{MeV})^3 \;\;,  \nonumber\\
&& A=\lambda (87.4\;\mbox{MeV})^2 \;\;.
\end{eqnarray}

  The parameter $A$ depends on temperature. It vanishes at
$T_{ch}=160 - 200\mbox{MeV}$. If $H=0$, $T_{ch}$ would be the temperature
of the second order chiral symmetry breaking phase transition. For $H\neq 0$
there is no phase transition but
a crossover; $\sigma_0(T)=< \sigma >_T$ is nonzero and it varies smoothly
with temperature, see Fig.1. The pion and sigma masses can be extracted from
small fluctuations around this minimum, see Fig.2. The masses remain
substantially nonzero nearby $T_{ch}$. Correlation lengths remain finite
and there is no critical slowing down.

   We assume that $T_{conf}\approx T_{ch}$. The initial conditions at
$T_{ch}$ are given by thermal equilibrium distributions. These initial
conditions themselves rule out any possibility of baryon production.

   The effective potential at
$T_{ch}=160\mbox{MeV}$ where $A(T_{ch})=0$ is shown in Fig.3. It has a unique
minimum at $\sigma_{0}(T_{ch})=44\mbox{MeV}$, the sigma mass is
$M_{\sigma}(T_{ch})=340\mbox{MeV}$. The correlation length of the $\sigma$
field is given by $\xi_{\sigma}=1/M_{\sigma}(T_{ch})\approx 0.6\mbox{fm}$, this is
roughly the size of the correlated domain of this field. The mass of the pion
field is $M_{\pi}(T_{ch})=196\mbox{MeV}$, its correlation length is
$\xi_{\pi}=1/M_{\pi}(T_{ch})=1\mbox{fm}$. We define the field
$\bar{\sigma}$, which is an average of $\sigma$ over its correlation domain.
In our calculations the average is defined by a cut-off in momentum space
at $M_{\sigma}$. Fluctuation squared of $\bar{\sigma}$ around its average is 
given by

\begin{eqnarray}\label{fluctuation}
\mu^2 &=& <[\bar{\sigma}-\sigma_{0}(T_{ch})]^2>= \nonumber\\
      &=& \int_{0}^{M_{\sigma}(T_{ch})}
          \frac{ 4\pi k^2 dk }{ (2\pi)^3 }
          \frac{ T_{ch} }{ M_{\sigma}^{2} + k^2 }=    \nonumber\\
      &=&\frac{T_{ch}M_{\sigma}(T_{ch})}{2\pi^2} (1-\frac{\pi}{4}) \;\;.
\end{eqnarray}
A gaussian distribution $f(\bar{\sigma};\mu)$ 
of these domain fluctuations is shown on an insert
in Fig.3. The same figure shows a cross section through the potential at zero
temperature. As the temperature drops below $T_{conf}\approx T_{ch}$ there
is no quark-gluon plasma to provide an external heat-bath for the chiral
fields. They evolve freely in the zero temperature potential from the
initial conditions given by thermal equilibrium distributions at $T_{ch}$.  

   From a topological point of view the skyrmion is a map from the space
compactified to $S^3$ to the $S^3$ bottom of the zero temperature sombrero
potential. As the vector field wraps around the target sphere, there must be
such a point that the field points in the $(\sigma,\vec{\pi})=(-1,\vec{0})$
direction. The position of a skyrmion can be identified with this point.
At $T_{ch}$ there are $\xi_{\sigma}$-sized domains of $\sigma$ and
$\xi_{\pi}$-sized domains of $\vec{\pi}$. The field at a given point will
roll down to the bottom of the sombrero potential in the $(-1,\vec{0})$
direction, if the domain averaged $\bar{\vec{\pi}}=0$ and at the same
time the domain averaged $\bar{\sigma}$ is to the left of the top of
the zero temperature potential, $\bar{\sigma}<-11.2\mbox{MeV}$.
The second condition is satisfied with the probability
\begin{equation} 
P[\bar{\sigma}<-11.2\mbox{MeV}]=\int_{-\infty}^{-11.2} f(\bar{\sigma };\mu ) 
d\bar{\sigma} = 0.01, 
\end{equation}
where $f(\bar{\sigma };\mu )$ is the gaussian distribution with fluctuation
given by (\ref{fluctuation}). The density of the points such that
the domain averaged $\bar{\vec{\pi}}=0$ can be worked out with
the general formula \cite{halperin,liu}

\begin{eqnarray}\label{halperin}
N[\bar{\vec{\pi}}=0]&=&
\frac{1}{\pi^2}
\{\frac{ < -\bar{\vec{\pi}}(\vec{x}) \; \nabla^2 \; \bar{\vec{\pi}}(\vec{x}) > }
       { < \bar{\vec{\pi}}(\vec{x}) \; \bar{\vec{\pi}}(\vec{x}) > }\}^{3/2}=
\nonumber \\
&=&
\frac{1}{\pi^2}
\{\;\; 
\frac{ \int_{0}^{M_{\pi}} k^2 dk \frac{k^2}{ M_{\pi}+k^2 }   }
     { \int_{0}^{M_{\pi}} k^2 dk \frac{ 1 }{ M_{\pi}+k^2 }   }  \;\;\}^{3/2}=
\nonumber \\
&=&
\frac{ M_{\pi}^3 }{\pi^2} \{ \frac{3\pi-8}{12-3\pi} \}^{3/2}\approx
\frac{0.04 }{\xi_{\pi}^3}  \;\;,
\end{eqnarray}
where $M_{\pi}=M_{\pi}(T_{ch})$ and $\xi_{\pi}$ is the pion correlation
length at $T_{ch}$. The ratio of the second derivative of the correlation
function to the correlation function itself is a proper measure of the density
of zeros. Indeed if the order parameter takes non-zero value and oscillates with the
small amplitude the correlation function is large in compare to
its second derivative and as a result its ratio is small. In the
case when the order parameter oscillates near zero with small frequency 
the correlation function is
comparable with its second derivative and density of zeros is of order unity
while for large frequency the second derivative of the correlation function is
larger then the function itself and density of zeros is substantial. The power
$3/2$ is \ref{halperin} for dimensional reasons.

As our definition does not distinguish baryons
from antibaryons, the expected density of antibaryons is

\begin{eqnarray}\label{density}
N_{\bar{B}} &=& \frac{1}{2}\times
                P[\bar{\sigma}<-11.2\mbox{MeV}]\times
                N[\bar{\vec{\pi}}=0]=  \nonumber\\
            &=& 2\cdot 10^{-4}\mbox{antibaryons}/\mbox{fm}^3 \;\;
\end{eqnarray}
for the assumed $T_{ch}=160\mbox{MeV}$.

\subsection*{Discussion}

We found negligible (anti-)baryons density (of order 
$10^{-4}\mbox{antibaryons}/\mbox{fm}^3$) after the chiral symmetry phase transition 
in quark-gluon plasma at zero baryon chemical potential. The (anti-)baryon density 
suppression results from the initial condition of the order parameter which is in 
thermal equilibrium at the critical point
and from the non zero light quark masses which substantially bias these
initial conditions. The probability of the 
topological defect formation is exponentialy supressed by the explicit
symmetry breaking parameter present in the free energy describing the
phase transition. This is the main source of discrepancy between our
prediction and that of Ellis, Kowalski and Heinz \cite{ellis2}.

  We assume that the chiral fields are 
relevant degrees of freedom in the vicinity
of critical temperature and that the transition is a smooth crossover. 
We expect that even in the case of the first order phase transition (resulting
from the strange quark mass and/or nonzero baryon chemical potential)
symmetry breaking bias would be strong enough to suppress baryon production.
This issue deserves a more careful investigation.

  Our result may overestimate the (anti-)baryon density.
As the chiral fields cool down the skyrmions can unwind through the top
of the sombrero potential. This process lasts until the system cools down 
to its Ginzburg temperature and can vastly decimate the defects.
This further reduction takes place if unwinding is not forbiden.
However, according to Holzwarth the unwinding must be forbiden for
the sake of baryon number conservation. In his model \cite{holzwarth}
the $(\sigma,\vec{\pi})=(0,\vec{0})$ point is removed from the configuration
space of the linear sigma model. In this case our estimate is accurate.

  The result of our calculation leads us to the conclusion that hardly any 
baryons will be produced in the central rapidity region of a heavy ion 
collision.

{\bf Acknowledgements}

We would like to thank A. Bialas, W. Czyz, G. Holzwarth, K. Zalewski and W.H. Zurek for 
stimulating discussions. J.D. was supported by the KBN grant 2 P03B 008 15 and
M.S. was supported by the KBN grant 2 P03B 086 14.

\centerline{\epsfbox{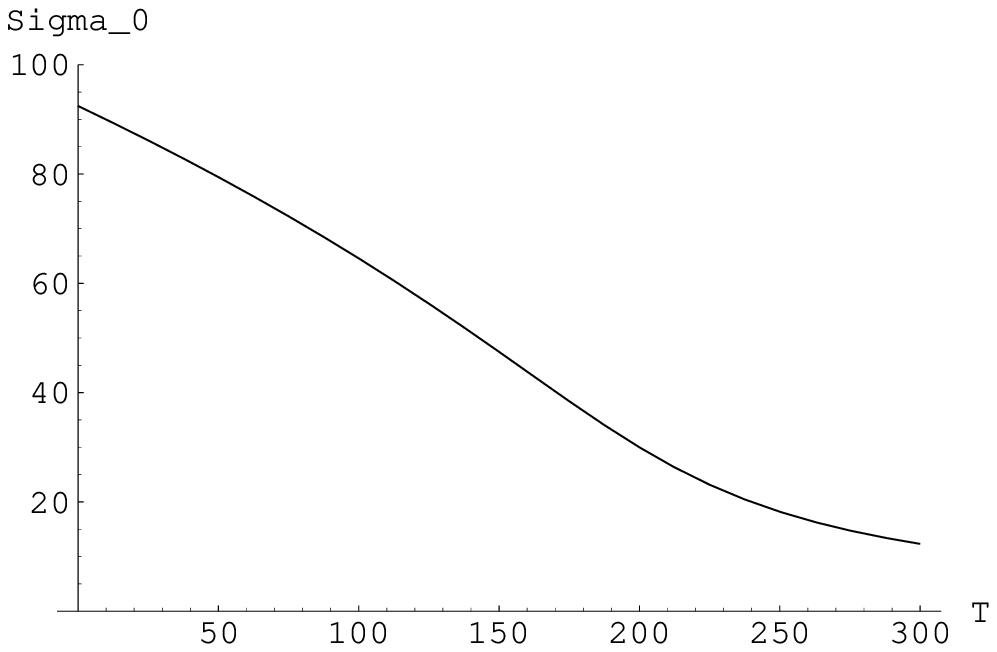}}
{\bf Fig.1. $\sigma_{0}(T)=<\sigma>_T$ [MeV] as a function of temperature
 [MeV] for the linear $\sigma$ model with the effective
 $A_{eff}(T)=A(1-T/T_{ch})$ and $T_{ch}=160\mbox{MeV}$. }

\centerline{\epsfbox{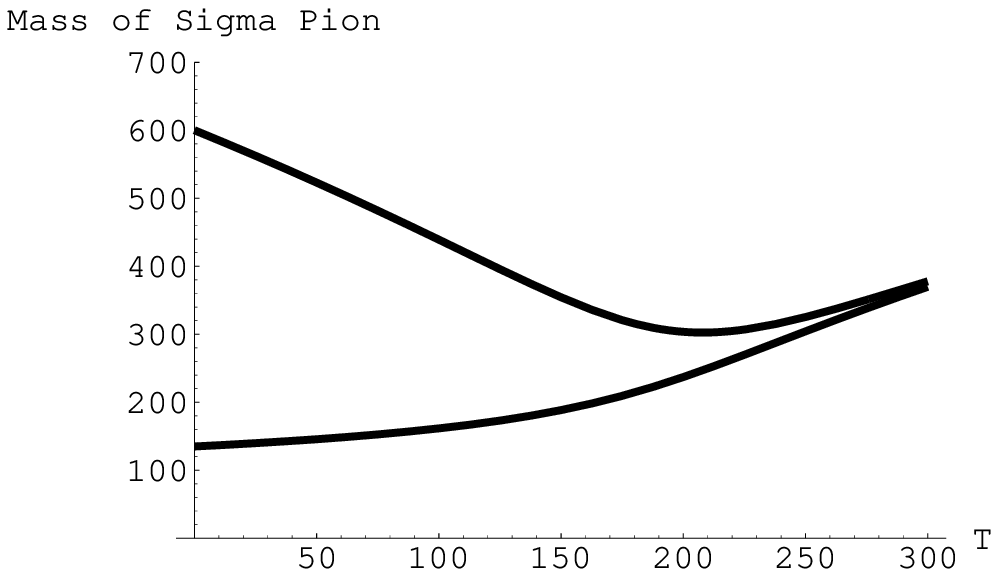}}
{\bf Fig.2. Temperature [MeV] dependence of the $\sigma$ (top) and $\pi$ (bottom) 
masses [MeV] under the same model assumptions as in Fig.1.}

\centerline{\epsfbox{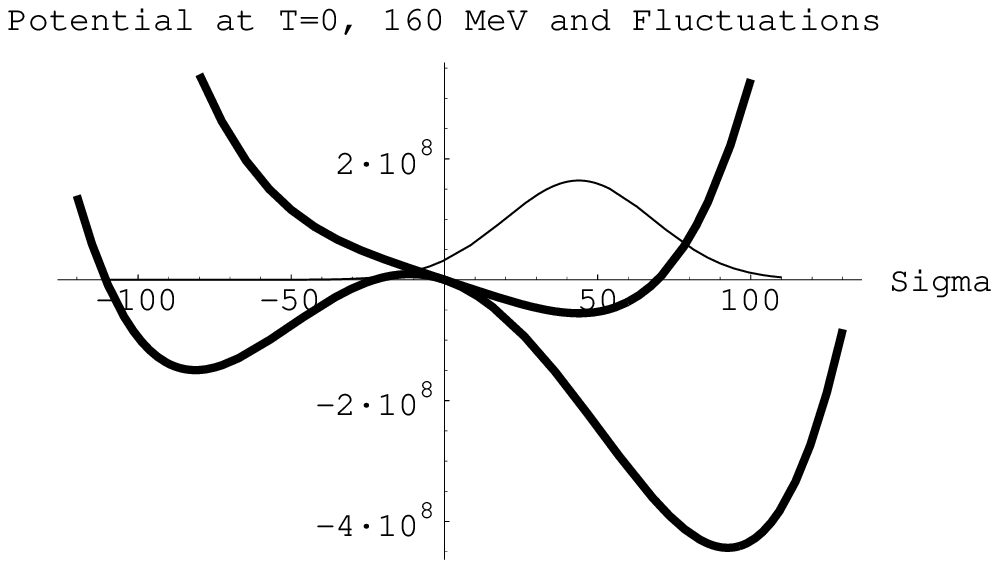}}
{\bf Fig.3. The effective potential $V_{eff}(\sigma,\vec{0})$ [MeV$^4$]
as a function of $\sigma$ [MeV]
at $T=T_{ch}$ (top) and at $T=0$ (bottom) - bold lines.
The gaussian distribution of $\bar{\sigma}$ at $T_{ch}$  - thin line. }

\end{document}